\documentclass{emulateapj}

\newcommand{\ltsimeq}{\raisebox{-0.6ex}{$\,\stackrel 
        {\raisebox{-.2ex}{$\textstyle <$}}{\sim}\,$}}

\newcommand{\civ}{C\,{\sc iv}}
\newcommand{\lya}{Ly\,$\alpha$}

\newcommand{\halpha}{H\,$\alpha$}
\newcommand{\oiii}{[O\,{\sc iii}]}

\newcommand{\asec}{^{\prime\prime}}

\def\co21{CO\,(2-1)}

\shorttitle{Molecular gas observations of the reddened quasar 3C\,318}
\shortauthors{Willott et al.}

\begin{document}


\title{Molecular gas observations of the reddened quasar 3C\,318}


\author{
Chris J. Willott\altaffilmark{1,2},
Alejo Mart\'inez-Sansigre\altaffilmark{3,4},
Steve Rawlings\altaffilmark{4},
}

\altaffiltext{1}{Herzberg Institute of Astrophysics, National Research
Council, 5071 West Saanich Rd, Victoria, BC V9E 2E7, Canada}
\altaffiltext{2}{Present address: University of Ottawa, Physics Department,
150 Louis Pasteur, MacDonald Hall, Ottawa, ON K1N 6N5,  Canada; cwillott@uottawa.ca}
\altaffiltext{3}{Present address: Max Planck Institute for Astronomy, koningstuhl 17, D-69117, Heidelberg, Germany; martinez@mpia-hd.mpg.de}
\altaffiltext{4}{Astrophysics, Department of Physics, Keble Road, Oxford, OX1
3RH, UK; sr@astro.ox.ac.uk}

\begin{abstract}
3C\,318 is a $z=1.574$ radio-loud quasar. The small physical size of
its radio jets indicate that these jets were triggered relatively
recently. In addition to the ultraviolet continuum emission being
reddened by dust, detections with {\it IRAS} and SCUBA show it to have
an exceptionally high far-infrared luminosity. We present \co21
observations of 3C\,318 made with the IRAM Plateau de Bure
Interferometer. We detect \co21 emission with a FWHM$\,=200\,{\rm
km\,s}^{-1}$ at a signal-to-noise ratio of $5.4$. There is evidence 
for positional ($\sim 20$\,kpc) and
velocity ($\sim -400\,{\rm km\,s}^{-1}$) offsets between the molecular
gas and the quasar which may be due to the quasar experiencing a major
merger. The mass of molecular gas inferred from our observations
is $M_{\rm H_2}=(3.0 \pm 0.6) \times 10^{10}\,M_{\sun}$. This
molecular gas mass is comparable to that in sub-mm-selected galaxies
at similar redshifts. The large molecular gas mass is consistent with
the primary source of heating for the cool dust in this quasar to be
massive star formation with a star formation rate of 
$1700\,M_{\sun}\,{\rm yr}^{-1}$ and a gas depletion timescale of
20\,Myr. Our observations support the idea that star formation
episodes and jet triggering can be synchronised.
\end{abstract}

\keywords{quasars:individual (3C\,318)---radio lines:$\>$galaxies---galaxies:starburst}

\section{Introduction}

Quasars are believed to be actively accreting black holes hosted by
massive galaxies. The discovery of supermassive black holes in the
nuclei of all nearby galaxies with stellar bulges (Magorrian et
al. 1998) indicates that such activity is not exceptional but the
norm. What is still unknown is exactly what physical processes are
responsible for the black hole mass -- stellar bulge mass correlation.

Of particular interest is to observe the simultaneous growth of black
holes and stellar mass in galaxies at the peak epoch of activity
(redshifts $1<z<3$). However, surveys of sources detected in hard
X-rays (probing AGN activity) and sub-millimetre emission (probing
star formation) found very little overlap amongst the brightest
sources in each class (Almaini et al. 2003; Waskett et al
2003). Only a low level of black hole accretion is found for most
sub-millimetre-selected galaxies (SMGs; Alexander et al. 2005).
Therefore the most intense periods of star formation and black
hole growth are typically decoupled. However, to be consistent with
the black hole mass -- stellar bulge mass correlation, models have
been proposed whereby the X-ray bright and sub-mm bright phases are
part of an evolutionary sequence within a single event, such as a
major merger (Granato et al. 2001; Archibald et al. 2002).

Radio-loud quasars possess a particularly useful feature. Their
relativistic jets can be modelled to infer the age of the radio source
and hence the time when the jet was triggered. Recent multi-wavelength
observations have revealed several interesting properties of {\em young}
radio sources: (i) strong associated \civ\ absorption (Baker et
al. 2002); (ii) high sub-mm luminosities (Willott et al. 2002); (iii)
fast, massive outflows of neutral gas (Morganti et al. 2005).  These
observations point towards a phase of great activity in the host
galaxies of sources with recently triggered radio jets.

A long standing issue in relation to the sub-mm detections of luminous
quasars is whether the radiating dust is heated by young stars or the
active nucleus (Almaini et al. 1999). The correlation between quasar
optical luminosity and sub-mm luminosity (Willott et al. 2003) could
be caused by direct heating of the dust by the active nucleus, or it
could instead be due to correlations between black hole and galaxy
mass (Omont et al. 2003). A significant number of sub-mm--luminous
quasars have been found to possess large reservoirs of molecular gas
via CO line observations (see Solomon \& Vanden Bout 2005 for a
compilation). Although time consuming work, such observations are
critical because molecular gas provides the fuel for star formation.
The existence of a large reservoir of molecular gas
provide useful insights on the evolutionary state of the system via its 
gas-to-dust ratio.

There are now 16 $z>1$ quasars with measured CO emission of which at
least 11 are known to have their CO detections assisted by
gravitational lensing. However, only one of these quasars, Q0957+561,
lies at redshifts $1<z<2$ when the bulk of the accretion in optical
quasars occurred (Barger et al. 2005). This redshift range contains
very few known CO emitters with the only other ones being the
extremely red starburst galaxy HR10 (Andreani et al. 2000) and a
gravitationally-lensed SMG SMM\,J02396 (Greve et al. 2005). It is also
worthwhile to note that most CO line observations of high-redshift
objects are of the high-$J$ transitions (Solomon \& Vanden Bout
2005). The low-$J$ transitions are less biased by the details of the
excitation and are more likely to offer a reliable estimate of the
total molecular gas mass (Papadopoulos \& Ivison 2002; Hainline et
al. 2006).

We have used the IRAM Plateau de Bure Interferometer (PdBI) to search
for low-$J$ CO emission in the $z=1.574$ radio-loud quasar 3C\,318. As
discussed in Willott et al. (2000), 3C\,318 is interesting because (i)
it is one of the most distant far-IR sources detected with {\it IRAS}
and therefore has a very high far-IR luminosity, (ii) it has a high
sub-mm luminosity measured with SCUBA, (iii) the quasar ultraviolet
continuum is reddened by dust and (iv) it is a very compact
(i.e. young) radio source. Here we report the discovery of \co21
emission in this quasar and discuss the implications. Cosmological
parameters of $H_0=71~ {\rm km~s^{-1}~Mpc^{-1}}$, $\Omega_{\mathrm
M}=0.26$ and $\Omega_\Lambda=0.74$ (Spergel et al. 2006) are assumed
throughout.

\section{Observations \& Results}

3C\,318 was observed in the 3 mm band with the PdBI, in the 6D (6
antennae, compact) configuration, on the nights of 2002 April 19 and
22. The weather conditions were generally excellent, particularly the
seeing ($0.74\asec$ and $1.01\asec$ on April 19 and 22 respectively)
leading to very good phase stability. The central frequency was tuned
to 89.60\,GHz, close to 89.5641\,GHz which is the frequency of the \co21
transition at a redshift of $z=1.574$ (the redshift of the narrow
\oiii\ emission line detected in near-IR spectroscopy by Willott et
al. 2000). The maximum bandwidth of 560 MHz was used, equivalent to
$1875\,{\rm km\,s}^{-1}$.

The observations lasted a total of 13.4 hours, which included
observations of the calibrators 3C\,273 (beginning of the night) and
MWC\,349 (end of the night) as well as regular visits to the phase
calibrator 1413+135. The data were reduced at IRAM in Grenoble. After
amplitude and phase calibrations, 3C\,318 was detected very strongly at
3\,mm. Due to incomplete coverage of the $uv$ plane, the beam includes
two artificial sidelobes, although this is not important for a
detection experiment and 3C\,318 appears unresolved anyway.

The continuum flux density of 28.65$\pm$0.28\,mJy is consistent with a
point source convolved with the beam (of size $8.05\asec \times
4.32\asec$), with the location of the peak just $0.2\asec$ from the
position of the compact radio source.  3C\,318 has been detected at
1.25\,mm with MAMBO at the IRAM 30-m Telescope and the relative fluxes
of the 1.25\,mm and 850\,$\mu$m emission indicate that the 1.25\,mm
emission is dominated by non-thermal synchrotron (Haas et
al. 2006). Therefore, the 3\,mm continuum must be entirely due to
synchrotron from the compact radio source.

A search of the data-cube using a routine kindly provided by Roberto
Neri revealed a fairly broad emission line close to the location of
the continuum emission. This was the only significant ($5\sigma$)
broad peak to be found in the entire field-of-view, giving
confidence that it is not spurious. The continuum flux density
contribution was then subtracted from the spectrum (in the $uv$ plane),
assuming a flat spectrum through the band. A frequency window was
previously specified around the line, however, so the flux of the
continuum to be subtracted did not include the line. After this
continuum subtraction, a spectrum was extracted from the data-cube,
using a polygon matching the shape of the synthesised beam.

\begin{figure}
\resizebox{0.45\textwidth}{!}{\includegraphics{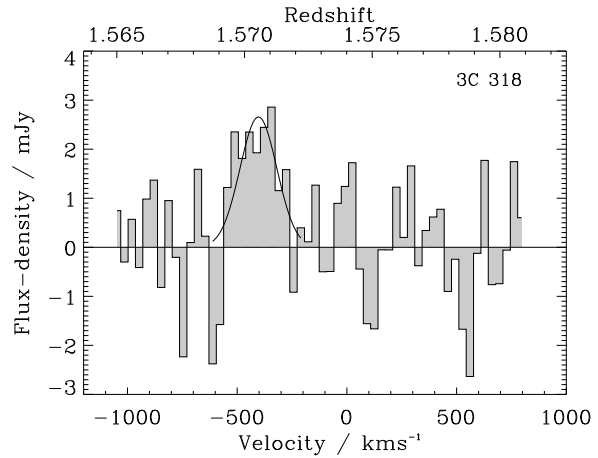}}
\caption{Spectrum of \co21 emission in 3C\,318. The data have been
rebinned into 20\,MHz ($33\,{\rm km\,s}^{-1}$) channels. The curve
shows the result of a single Gaussian fit to the line with a FWHM of
${\rm 200\,km\,s}^{-1}$. The upper axis shows the redshift
corresponding to the \co21 transition.
\label{fig:spec}
}
\end{figure}

The spectrum was rebinned to 20\,MHz ($33\,{\rm km\,s}^{-1}$) channels and is
shown in Fig. 1. Positive emission is visible in nine adjacent
channels, corresponding to a full width zero intensity (FWZI) of
$300\,{\rm km\,s}^{-1}$. Integrating the emission over these channels
results in a line flux of $1.19 \pm 0.22\,{\rm Jy\,km\,s}^{-1}$. Also shown
in Fig.\,\ref{fig:spec} is a Gaussian fit to the emission line. The
best fit Gaussian has a centre at $-403\,{\rm km\,s}^{-1}$ and a FWHM
of ${\rm 200\,km\,s}^{-1}$. The emission is somewhat more boxy than a
Gaussian, possibly indicating that a multiple Gaussian fit may
describe the kinematics better, as has been found for several SMGs
(Greve et al. 2005).  However, our data only have a signal-to-noise
ratio for the line of 5.4, so a multi-component fit would not be well
constrained and is not attempted. 

\begin{figure}
\resizebox{0.45\textwidth}{!}{\includegraphics{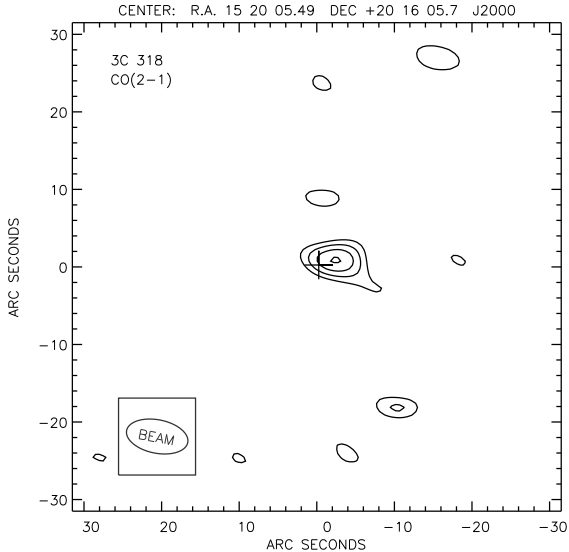}}
\caption{The contours show the \co21 emission from 3C\,318 (integrated
  over channels corresponding to $-550\,{\rm km\,s}^{-1}$ to
  $-270\,{\rm km\,s}^{-1}$). The location of the 3\,mm continuum peak
  (dominated by synchrotron emission from the compact radio source) is
  indicated with a cross. The offset between these positions is
  significant at the $4\sigma$ level. The beam size is $8.05\asec \times
  4.32\asec$ and is shown in the box in the lower-left corner. Contours are drawn at 0.7, 
1.05, 1.40, 1.75 mJy\,beam$^{-1}$.
\label{fig:map}
}
\end{figure}

A map of the line emission was made by extracting the velocity
channels containing the line. This map is contoured in
Fig.\,\ref{fig:map}. The peak of the line emission is $2.4\asec$ west
and $0.5\asec$ north of the continuum peak. We now consider the
significance of this offset, given the relatively low signal-to-noise
ratio (SNR) of the emission line. The beam shape in the map is an
ellipse with Gaussian FWHM of $8.05\asec$ and $4.32\asec$ at a
position angle of $79^{\circ}$ east of north.  In the one-dimensional
case, the $1\sigma$ positional uncertainty is given by $\sigma_{x} = ({\rm
FWHM}/2)/{\rm SNR}$ (e.g. Downes et al. 1999). For the two axes of the
elliptical beam we find $\sigma_{x}=0.75\asec$ and
$\sigma_{y}=0.40\asec$. Along the angle corresponding to the offset of
$2.4\asec$ west and $0.5\asec$ north, the size of the $1\sigma$
positional uncertainty is $0.65\asec$ and hence the significance of
the positional offset is $4\sigma$. We note that given the low SNR
and resolution of the line, we cannot rule out a considerable fraction of
the molecular gas existing at the same site as the synchrotron continuum.

In addition to this spatial offset, the velocity peak of the molecular
line emission is blueshifted by $400\,{\rm km\,s}^{-1}$ compared to
the redshift determined from the narrow \oiii\ emission. The narrow emission
lines of a quasar are usually a good indicator of the galaxy systemic
velocity (Vrtilek \& Carleton 1985). Greve et al. (2005) consider the
velocity offsets between molecular and ultraviolet/optical emission lines in
quasars and SMGs. They find that offsets of several hundred ${\rm
km\,s}^{-1}$ are common for the molecular and \lya\ lines for quasars
and SMGs. However, they find very small offsets ($<100\,{\rm
km\,s}^{-1}$) for the molecular and \halpha\ lines in four SMGs. The
explanation for this is that in SMGs, \halpha\ traces the same
star-forming regions that produce the molecular emission and is not so
affected by dust obscuration and absorption as \lya.  For the case of
quasars it is well-known that there are large offsets between the
broad and narrow emission lines (e.g. Richards et al. 2002) and that
it is difficult to locate the true \lya\ peak at high redshift. 

How should we interpret the projected spatial ($\sim 20$\,kpc) and
velocity ($\sim 400\,{\rm km\,s}^{-1}$) offsets between the radio core
(situated close to the supermassive black hole at the bottom of the
galaxy potential well) and the bulk of the molecular gas? We note that
similar sized offsets have been found for SMGs between the peak of the
molecular gas and the peak of the $K$-band stellar continuum (Tacconi
et al. 2006). The most likely explanation is that 3C\,318 is in the
early stages of a major merger, when such offsets between the two
galaxies are common (Patton et al. 2000). In this case, we find that
the bulk of the molecular gas (and star formation) is in the companion
galaxy to the active radio source.  Higher resolution observations of
the molecular gas and thermal dust emission are clearly needed to
fully understand the situation in this complex system.

We calculate the CO line luminosity and molecular gas mass assuming a
conversion factor $\alpha=0.8({\rm K\,km\,s}^{-1}\,{\rm
pc}^{2})^{-1}\,M_{\sun}$, as in Solomon \& Vanden Bout (2005). The
line luminosity is $L'_{\rm CO}=(3.8 \pm 0.7) \times 10^{10}\,{\rm
K\,km\,s}^{-1}\,{\rm pc}^{2}$ and the mass of molecular hydrogen
estimated to be $M_{\rm H_2}=(3.0 \pm 0.6) \times
10^{10}\,M_{\sun}$. These values are comparable with the mean of a
sample of CO-detected SMGs (Greve et al. 2005). We note that there is
no evidence for gravitational lensing of 3C\,318 (Willott et
al. 2000).

\begin{figure}
\resizebox{0.45\textwidth}{!}{\includegraphics{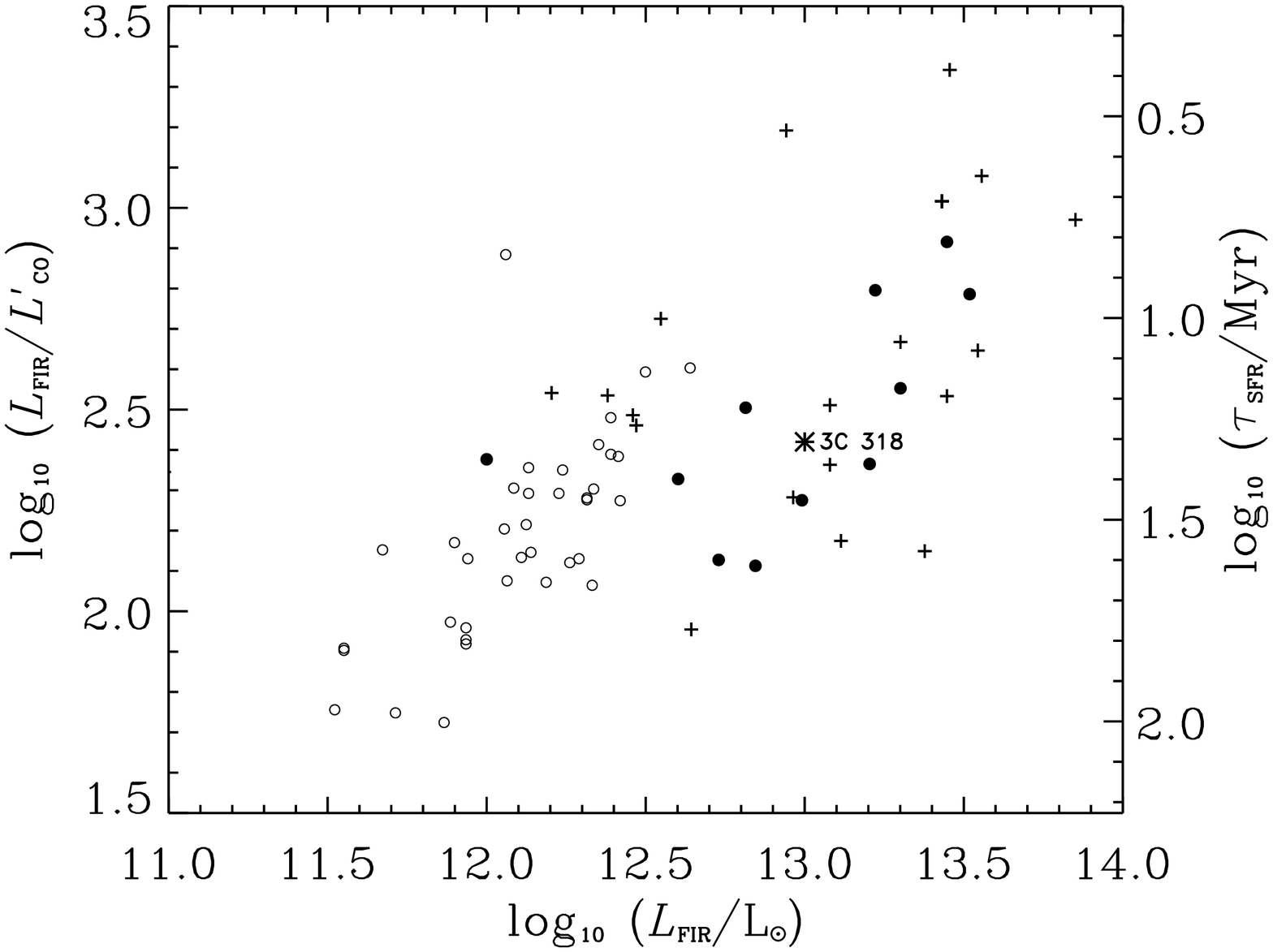}}
\caption{Plot of $L_{\rm FIR}/L'_{\rm CO}$ ($\propto$ SFR timescale)
vs $L_{\rm FIR}$ for ULIRGS (open circles; Solomon et al. 1997), SMGs
(filled circles; Solomon \& Vanden Bout 2005), quasars and radio
galaxies (plus signs; Solomon \& Vanden Bout 2005) and 3C\,318 (asterisk). The
right axis shows the molecular gas depletion timescale assuming a star
formation rate derived from the far-infrared luminosity.
\label{fig:time}
}
\end{figure}

To compare the mass of molecular gas with the rate at which
it is being turned into stars we use the far-infrared luminosity to
estimate the star formation rate.  Willott et al. (2000) showed that
3C\,318 has an extremely high far-infrared luminosity with significant
contributions from dust warmer than that which is typical of
SMGs. Since this warm dust component is likely heated by the
quasar rather than young stars, we calculate the far-infrared
luminosity using only the SCUBA $850\mu$m flux-density and assuming a
cool ($T=41$K) dust spectrum as in Willott et al. (2002), yielding the
relation $L_{\rm FIR} \approx 2 \times 10^{12}\,(S_{850}\,/\,{\rm
mJy})\,L_{\sun}$. The measured $850\mu$m flux-density of 3C\,318 is
$7.78 \pm 1.00$\,mJy (Willott et al. 2002). Haas et al. (2006) detect
3C\,318 with MAMBO at 1.25\,mm with a flux-density of
$5.61\pm1.42$\,mJy. This suggests some fraction of the $850\,\mu$m
flux is due to synchrotron and from consideration of the
synchrotron spectral slope we estimate the thermal contribution at
$850\,\mu$m to be $\approx 5$\,mJy. Therefore the cool dust
far-infrared luminosity is $\log_{10} L_{\rm FIR} = 13.0\,L_\sun$.
Integrating over the stellar initial mass function, this corresponds
to a star formation rate of $1700\,M_{\sun}\,{\rm yr}^{-1}$ (Kennicutt
1998) and a corresponding timescale for molecular gas depletion of
$\sim 20$\,Myr.  By comparison, the compact radio jets extend only
7\,kpc and, assuming a hotspot advance speed of $\leq 0.1c$,
were triggered $\ltsimeq 1$\,Myr before the time of observation.


In Fig.\,\ref{fig:time}, we show how these values for the far-infrared
luminosity, ratio of far-infrared to CO luminosity and molecular gas
depletion timescale compare with all known $z>1$ molecular line
emitters from the literature (Solomon \& Vanden Bout 2005) and
low-redshift ULIRGs (Solomon et al. 1997). This plot shows that
high-redshift quasars and radio galaxies occupy a similar location to
SMGs and that 3C\,318 is typical of both of these populations.

The discovery of a large molecular gas mass in 3C\,318 has
implications for the power source of the exceptionally luminous dust
emission. In the case of dust heated directly by the active nucleus, a
large molecular gas mass is not required. The measured
molecular gas mass is equivalent to that of the typical SMG which has
only a weak active nucleus (Alexander et al. 2005). Therefore, our
observations suggest that the 3C\,318 system is undergoing a starburst
comparable to that of the typical SMG.

\section{Conclusions}

We have carried out a sensitive search for \co21 emission in the
$z=1.574$ quasar 3C\,318. We find a molecular gas mass of $M_{\rm
H_2}=(3.0 \pm 0.6) \times 10^{10}\,M_{\sun}$. This value is typical
for SMGs, leading to the conclusion that the predominant power source
for the exceptionally high cool dust luminosity in this quasar is due to
star formation. 

The implications of these observations go beyond this one quasar and
provide extra evidence towards the synchronisation of star formation
and jet formation events in some radio-loud AGN. Willott et al. (2002)
found that small (young) radio sources, such as 3C\,318, are more
likely to be strong sub-mm sources than large (old) radio
sources. Additionally, several of the most sub-mm-luminous quasars in
Willott et al. (2002) have reddened optical spectra (like 3C\,318),
indicating significant dust in their host galaxies. There is also a
tendency for the most sub-mm-luminous quasars to have strong \civ\
associated absorption. Baker et al. (2002) presented a negative
correlation between \civ\ associated absorption equivalent width and
radio source size (age). Morganti et al. (2005) found that small radio
sources have fast ($\sim 1000\,{\rm km\,s}^{-1}$) outflows carrying a
considerable mass of neutral gas. Putting all these lines of evidence
together points towards a dusty, metal-enriched interstellar medium
with a high rate of star formation in quasars with recently triggered
radio jets. The larger-scale massive gas outflows which follow may be
effective at clearing out much of this material and suppressing star
formation. As shown by Morganti et al., such outflows could be
energetic enough to not only inhibit star formation, but expel the
remaining gas from the galaxy. Therefore, radio sources may be
significant contributors to the feedback required to explain the
observable properties of elliptical galaxies and their black holes.

\acknowledgments Based on observations carried out with the IRAM
Plateau de Bure Interferometer. IRAM is supported by INSU/CNRS
(France), MPG (Germany) and IGN (Spain). Thanks to the IRAM staff,
particularly Helmut Wiesemeyer for his extensive (and patient) help
with the data reduction, and Roberto Neri for use of his
software. Thanks to the anonymous referee for many useful suggestions.

\end{document}